\documentclass[12pt]{iopart}

\usepackage{iopams,graphicx,amssymb}
\eqnobysec

\begin{document}

\title[Scaling in erosion of landscapes]
{Scaling in erosion of landscapes:
Renormalization group analysis of a model with turbulent mixing}

\author{N. V. Antonov and P. I. Kakin}

\address{Department of High Energy and Elementary Particles Physics, St.~Petersburg State University,
Universitetskaya nab. 7-9., St.~Petersburg, 199034 Russia}

\ead{n.antonov@spbu.ru, p.kakin@spbu.ru}

\begin{abstract}
The model of landscape erosion, introduced in [{\it Phys. Rev. Lett.}  {\bf 80}: 4349 (1998); 
{\it J. Stat. Phys.} {\bf 93}: 477 (1998)] and modified in [{\it Theor. Math. Phys.} - in press; arXiv:1602.00432], is advected by anisotropic velocity field. The field is Gaussian with vanishing correlation time and the
pair correlation function of the form
$\propto \delta(t-t') / k_{\bot}^{d-1+\xi}$, where
$k_{\bot}=|{\bf k}_{\bot}|$ and ${\bf k}_{\bot}$ is the component of the
wave vector, perpendicular to a certain preferred direction -- the
$d$-dimensional generalization of the ensemble introduced
by Avellaneda and Majda [{\it Commun. Math. Phys.} {\bf 131}: 381 (1990)]. Analogous to the case without advection, the 
model is multiplicatively renormalizable and has infinitely many coupling constants. The one-loop counterterm
is derived in a closed form in terms of the certain function $V(h)$, entering the original 
stochastic equation, 
and its derivatives with respect to the height field $h(t,{\bf x})$. The full infinite
set of the one-loop renormalization constants, $\beta$-functions and anomalous dimensions is obtained from the Taylor expansion of the counter-term. 
Instead of a two-dimensional surface of fixed points there is two such surfaces; they are likely 
to contain infrared attractive region(s). If that is the case, the model exhibits scaling 
behaviour in the infrared range. The
corresponding critical exponents are nonuniversal because they depend
on the coordinates of the fixed points on the surface; they also satisfy certain universal exact relation.

\end{abstract}


\section{Introduction and description of the model} \label{sec:Intro}

The problem of landscape
erosion due to the flow of air or water over it and related
problems ({\it e.g.} granular flows) have been attracting constant interest over the past few decades; see Refs. \cite{E1}--\cite{E14}
and the literature cited therein. A plethora of widely varied physical phenomena is related to these issues which makes a construction of
underlying dynamical model a complicated task. These models have been a source of much controversy
\cite{E5}--\cite{E13}.
However, landscape erosion has some universal
aspects (like the exponents in scaling laws) that, in analogy with critical phenomena, can be
described within the framework of relatively simple semiphenomenological
models. Indeed, such models can be built on the basis of dimensionality and symmetry
considerations; see, {\it e.g.} the discussion in \cite{Pastor1,Pastor2} and
references therein.

Similar situation takes place in the related problem of kinetic
roughening of surfaces or interfaces, described by the well known
Kardar-Parisi-Zhang stochastic model \cite{KPZ}
and its descendants \cite{Kim}--\cite{AK1}. Another example is provided
by the problem of self-organized criticality, which is described in the continuum
limit by the Hwa-Kardar stochastic model \cite{HK}
and its modifications \cite{Tadic,AK2}. 

A model of landscape erosion of a surface with a fixed mean tilt was
proposed in \cite{Pastor1,Pastor2}. In \cite{US} it was shown that the proposed model is only renormalizable if it is modified to include the whole series in the powers of $h$ - the height of the propfile - instead of just its leading terms.

Let us describe the modified model.

A unit constant vector ${\bf n}$ determines a certain preferred
direction (tilt of the slope) and, therefore,
establishes an intrinsic anisotropy of the model.
Any vector can be decomposed into the components perpendicular and
parallel to ${\bf n}$. For the $d$-dimensional
horizontal position ${\bf x}$ one has
${\bf x} = {\bf x}_{\bot} + {\bf n} x_{\parallel}$ with
${\bf x}_{\bot} \cdot {\bf n} =0$.
In the following, we denote the derivative in the full $d$-dimensional
${\bf x}$ space by
$\partial = \partial/ \partial {x_i}$ with $i=1\dots d$, and the
derivative in the subspace orthogonal to ${\bf n}$ by
$\partial_{\bot}=\partial/ \partial {x_{\bot i}}$ with $i=1\dots d-1$.
Then the derivative in the parallel direction is written as 
$\partial_{\parallel} = {\bf n}  \cdot \partial$.

The stochastic differential equation for the height of the profile,
{\it i.e.} for the height field $h(x)=h(t,{\bf x})$, proposed in
\cite{Pastor1,Pastor2} and modified in \cite{US}, is taken in the form
\begin{equation}
\partial_{t} h= \nu_{\bot 0}\, \partial_{\bot}^{2} h + \nu_{\parallel 0}\,
\partial_{\parallel}^{2} h +
\partial_{\parallel}^{2} V(h) + f.
\label{eq1}
\end{equation}
Here $\partial_{t} = \partial/ \partial {t}$, $\nu_{\parallel 0}$ and
$\nu_{\bot 0}$ are topographic diffusion coefficients, $V(h)$ is some
function that depends only on the field $h(x)$ (and not on its derivatives)
and $f(x)$ is a Gaussian random noise with zero mean and prescribed
pair correlation function
\begin{equation}
\langle f(x)f(x') \rangle = D_0
\delta(t-t')\, \delta^{(d)}({\bf x}-{\bf x}')
\label{forceD}
\end{equation}
with some positive amplitude $D_0$. 

Here and below the subscript ``o'' means that the parameters in (\ref{eq1}) are bare, {\it i.e.} not yet renormalized.

The function $V(h)$ is a series in powers of $h(x)$. In \cite{Pastor1,Pastor2}
is was taken odd in $h$ which was explained by the
symmetry $h,f\to-h,-f$ (another symmetry of the model is
$x_{\parallel}\to-x_{\parallel}$). The authors of \cite{Pastor1,Pastor2} also
truncated the Taylor expansion of $V(h)$ on the leading $h^{3}$ term, but the whole series in $h$ should be considered instead \cite{US}.

When added to the problem, the various kinds of deterministic or chaotic
flows change behaviour
of the critical systems (like liquid crystals or binary mixtures near their
consolution points). Indeed, the flow can destroy the usual critical
behaviour, change it to the mean-field behaviour, or give rise to a new non-equilibrium
universality class \cite{Satten}--\cite{AIK}. That is why it is vital to study the influence of turbulent mixing on critical behavior. 

In this paper the velocity field will be modelled by the strongly anisotropic Gaussian ensemble, with vanishing
correlation time and prescribed power-like pair correlation function -- the $d$-dimensional generalization of the ensemble introduced and studied
in \cite{AM}. At the same time, the ensemle is an anisotropic modification of the
popular Kraichnan's rapid-change model; see \cite{FGV} for the review and
the references. Anisotropic flow is natural to consider because the
modified model of erosion already involves intrinsic anisotropy, related
to the overall tilt of the landscape. 

Coupling with the velocity field $v_{i}(x)$ is introduced by the replacement
\begin{equation}
\partial_{t} \to \nabla_{t} = \partial_{t} + v_{i} \partial_{i},
\label{nabla}
\end{equation}
where $\nabla_{t}$ is the Lagrangian (Galilean covariant) derivative.

The velocity field will be taken in the form
\begin{equation}
{\bf v} = {\bf n} v(t, {\bf x}_{\bot}),
\label{vello}
\end{equation}
where $v(t, {\bf x}_{\bot})$ is a scalar function
independent of $x_{\parallel}$. Then the incompressibility condition
is automatically satisfied:
\begin{equation}
\partial_{i} v_{i} = \partial_{\parallel} v(t, {\bf x}_{\bot}) = 0.
\label{inko}
\end{equation}
We assume that $v(t, {\bf x}_{\bot})$ has a Gaussian
distribution with zero mean and the pair correlation function of the form
\begin{eqnarray}
\langle v(t,{\bf x}_{\bot}) v(t',{\bf x}_{\bot}') \rangle = \delta(t-t')
\int
\frac{d k}{(2\pi)^{d}} \, \exp \left\{ {\rm i} k\cdot ({\bf x}-{\bf x}')
\right\}
D_{v} (k)= \nonumber \\
= \delta(t-t') \int \frac{d k_{\bot}}{(2\pi)^{d-1}} \, \exp
\left\{ {\rm i} k_{\bot}\cdot ({\bf x}_{\bot}-{\bf x}'_{\bot}) \right\}
\widetilde D_{v} (k_{\bot}),
\label{veloc1}
\end{eqnarray}
with $k_{\bot}=|k_{\bot}|$ and the scalar coefficient function $D_{v}$:
\begin{equation}
D_{v} (k)= 2\pi \delta(k_{\parallel}) \, \widetilde D_{v} (k_{\bot}) ,
\quad \widetilde D_{v} (k_{\bot}) = B_0\, k_{\bot}^{-d+1-\xi}.
\label{veloc2}
\end{equation}
Here $B_0 >0$ is a constant amplitude factor; $\xi$ is an arbitrary
exponent, which will play
the role of a formal RG expansion parameter (along with $\varepsilon=2-d$). The cutoff $k_{\bot}>m$ provides the infrared (IR)
regularization in (\ref{veloc1}). The precise form of the cutoff is unimportant; the sharp cutoff is the most convenient choice
from the calculational viewpoint.

We apply the standard field theoretic renormalization group (RG) to the modified model of erosion with turbulent mixing and arrive at the results similar to those presented in \cite{US}.

The plan of the paper and the main results are as follows.

In section \ref{sec:Model} the field theoretic formulation of
the stochastic problem (\ref{eq1}), (\ref{forceD}) for the arbitrary
(not necessarily odd) full-scale (not truncated) function $V(h)$ is presented.

In section \ref{sec:Reno} ultraviolet (UV) divergences and
renormalization procedure of the resulting field theory are discussed. The 
upper critical dimension is established to be $d=2$; this leads to the emergence of infinitely many coupling
constants in renormalized model, and, subsequently, to the emergence of infinitely many
$\beta$-functions in the corresponding RG equations.

We write down
the corresponding renormalized action functional, renormalization
relations for the fields and parameters, RG equations and RG functions
($\beta$-functions and anomalous dimensions).

In section~\ref{sec:Funk} the renormalization procedure is performed
in the leading one-loop order. Despite the fact that the model
involves infinitely many couplings, the one-loop counterterm
is derived in a closed form in terms of the function $V(h)$
and its derivatives with respect to the variable $h(x)$. Its Taylor expansion gives rise to the full infinite
set of one-loop renormalization constants, and, therefore, to all
$\beta$-functions and anomalous dimensions.

In this derivation, we adopt the functional method applied earlier by
A.~N.~Vasil'ev and one of the authors \cite{AV} to an isotropic model
of surface roughening, proposed in \cite{Pavlik} as a possible
modification of the Kardar-Parisi-Zhang equation; see also \cite{AA,AAA}. This method was also applied in \cite{US}.

In section~\ref{sec:Att} attractors of the obtained RG
equations are analyzed in the infinite-dimensional space of coupling constants.
Instead of a set of fixed points (like for most
multicoupling models), there are two two-dimensional surfaces of fixed points -- one of them corresponds to IR asymptotic regime, where turbulent mixing is irrelevant, and coincides with the surface obtained in \cite{US}. These surfaces are likely to contain IR attractive region(s).
If so, the model exhibits scaling behaviour in the IR range. The
corresponding critical exponents are nonuniversal because they depend on
the coordinates of the specific fixed points on the surface, but
satisfy certain exact relation.

The remaining problems are briefly discussed in section~\ref{sec:Conc}.

\section{Field Theoretic Formulation of the Model} \label{sec:Model}

According to the general statement (see, {\it e.g.} the books \cite{Zinn,Book3}
and the references therein), 
the stochastic problem
(\ref{eq1}), (\ref{forceD}) is equivalent to the field theoretic model
of the set of fields $\Phi=\{h,h',v\}$ with the action functional  
\begin{equation}
{\cal S}(\Phi)=h'h'+h'\left\{-\partial_{t}h+\nu_{\bot 0}\, \partial_{\bot}^{2} h + \nu_{\parallel 0}\,
\partial_{\parallel}^{2} h +
\partial_{\parallel}^{2}\sum_{n=2}^{\infty}\frac{\lambda_{n0} h^{n}}{n!}\right\}+{\cal S}_{\boldsymbol{v}}
\label{act1}
\end{equation}
(we have scaled out $D_0$ and other factors of $h'h'$ by adjusting the values of $\lambda_{n0}$).

The last term in (\ref{act1}) corresponds to the Gaussian averaging over
${\bf v}$ with correlator (\ref{veloc1}) and has the form
\begin{equation}
{\cal S}_{\boldsymbol{v}}= \frac{1}{2}\,
\int dt \int d{\bf x}_{\bot} d{\bf x}_{\bot}' v(t,{\bf x}_{\bot})
\widetilde D^{-1}_{v} ({\bf x}_{\bot}-{\bf x}'_{\bot}) v(t,{\bf x}_{\bot}'),
\label{Sv}
\end{equation}
where
\begin{equation}
\widetilde D^{-1}_{v} ({\bf r}_{\bot}) \propto B_0^{-1} \, r_{\bot}^{2(1-d)-\xi}
\label{Dv}
\end{equation}
is the kernel of the inverse linear operation $D^{-1}_{v}$ for the
correlation function $D_{v}$ in (\ref{veloc2}).

Here and below, all the needed integrations over $x = (t,{\bf x})$ are always implied, {\it e.g.}
\begin{equation}
h' h'=\int dt\int d{\bf x} \,\, h'(t,{\bf x})\, h'(t,{\bf x}).
\end{equation}

The field theoretic formulation of the stochastic problem identifies various correlation and
response functions of the stochastic problem (\ref{eq1}), (\ref{forceD})
with various Green's functions of the field theoretic model
with the action (\ref{act1}). In other words, the correlation fucntions are now represented by
functional averages over the full set of fields $\Phi=\{ h,h',v\}$ with
the weight $\exp {\cal S}(\Phi)$. 

A standard
Feynman diagrammatic technique applies to the model (\ref{act1}). There are three bare propagators (lines in the
diagrams): $\langle {\bf v}{\bf v} \rangle_{0}$ (given by (\ref{veloc1}),
(\ref{veloc2})), and the propagators of the scalar fields
(in the frequency--momentum and time--momentum representations):
\begin{equation}
\langle hh' \rangle_{0} = \langle h'h
\rangle_{0}^{*}
= \left\{-{\rm i} \omega+ \varepsilon(k)
\right\}^{-1}, \quad
\langle hh \rangle_{0} =  2 \,
\left\{ \omega^{2} + \varepsilon^{2}(k) \right\}^{-1},
\label{lines3}
\end{equation}
where $\varepsilon(k)=\nu_{\parallel}^2 k_{\parallel}^2+\nu_{\bot}^2
k_{\bot}^2$. The propagator $\langle h'h'\rangle_{0}$ vanishes identically
for any field theory of the type (\ref{act1}). The interaction terms
$-h'\partial_{\parallel} V(h)$ and $-h'(v\partial_{\parallel})h$ give rise to the vertices with bare coupling constants
$g_{n0}$ ($n=2,3,\dots$) and $w_0$:
\begin{equation}
\lambda_{n0}=g_{n0} \nu_{\parallel 0}^{(n+3)/4}\nu_{\bot 0}^{(n-1)/4}, \quad
B_0 = w_0\nu_{\parallel 0},
\label{D0}
\end{equation}
so that by dimension $g_{n0} \sim \ell^{-\varepsilon(n-1)/2}$ and
$w_0 \sim \ell^{-\xi}$,
where $\ell$ has the order of the smallest length scale in our problem.

\section{UV divergences and renormalization} \label{sec:Reno}

To analyze the UV divergences the analysis of canonical dimensions is used; see, {\it e.g.} \cite{Zinn,Book3}.
Dynamic models of the type (\ref{act1}) usually have two scales, {\it i.e.} their dimensions are described by   
the  two numbers - the frequency dimension $d_{F}^{\omega}$, and the momentum dimension $d_{F}^{k}$. 
These two numbers completely define the canonical dimension of a quantity $F$ (a field or a parameter): $[F] \sim [T]^{-d_{F}^{\omega}} [L]^{-d_{F}^{k}}$ ($L$ is the typical length
scale and $T$ is the time scale); see, {\it e.g.} Chap.~5 in book \cite{Book3}.
In the present case, there are two independent momentum scales because of the anisotropy of the model. Namely, two independent momentum canonical dimensions 
$d_{F}^{\bot}$ and $d_{F}^{\parallel}$ has to be introduced so that
\[ [F] \sim [T]^{-d_{F}^{\omega}}  [L_{\bot}]^{-d_{F}^{\bot}}
[L_{\parallel}]^{-d_{F}^{\parallel}}, \]
where $L_{\bot}$ and $L_{\parallel}$ are (independent) length scales in the
corresponding subspaces. The obvious
normalization conditions are $d_{k_{\bot}}^{\bot}= -d_{\bf x_{\bot}}^{\bot}=1$,
$d_{k_{\bot}}^{\parallel}=-d_{\bf x_{\bot}}^{\parallel}=0$,
$d_{k_{\bot}}^{\omega} = d_{k_{\parallel}}^{\omega}=0$,
$d_{\omega }^{\omega }=-d_t^{\omega }=1$, {\it etc.}; moreover, each term of the action functional (\ref{act1})
is assumed to be dimensionless with respect to all the three independent dimensions
separately. The original momentum dimension can be found from the
relation $d_{F}^{k} = d_{F}^{\bot}+ d_{F}^{\parallel}$.
Then, the
total canonical dimension is $d_{F}=d_{F}^{k}+2d_{F}^{\omega}  =
d_{F}^{\bot} + d_{F}^{\parallel} +2d_{F}^{\omega}$. The factor $2$ in the last term comes from the consideration that in the free theory
$\partial_{t}\propto\partial^{2}_{\bot} \propto \partial^{2}_{\parallel}$.

The canonical dimensions for the model (\ref{act1}) are presented in
table~\ref{table1}. The renormalized parameters (without the subscript ``o'') and the renormalization mass
$\mu$ will be introduced later.

\begin{table}[h]
\centering
\caption{Canonical dimensions of the fields and the parameters in the
model (\protect\ref{act1})} \label{table1}
\begin{tabular}{lllllllllll}
\hline
$F$ & $h'$ & $h$ & $v$ & $\nu_{\bot}$ & $\nu_{\parallel}$ & $\lambda_{n0}$ & $g_{n0}$ & $w_0$ & $g_n$, $w$ & $\mu$\\\hline
$d_{F}^{\omega}$ & $1/2$ & $-1/2$ & $1$ & $1$ & $1$ & $(n+1)/2$ & $0$ & $0$ & $0$ & $0$\\
$d_{F}^{\parallel}$ & $1/2$ & $1/2$ & $-1$ & $0$ & $-2$ & $-(n+3)/2$ & $0$ & $0$ & $0$&  $0$\\
$d_{F}^{\bot}$ & $(d-1)/2$ & $(d-1)/2$ & $0$ & $-2$ & $0$ & $(d-1)(1-n)/2$ & 
$(2-d) (n-1)/2$ & $\xi$ & $0$ & $1$\\
\hline
$d_{F}$ & $d/2+1$ & $-(2-d)/2$ & $1$ & $0$ & $0$ & $(2-d)(n-1)/2$ 
& $(2-d) (n-1)/2$ & $\xi$ & $0$ & $1$\\
\hline
\end{tabular}
\end{table}

As could be seen from table~\ref{table1}, all the coupling constants $g_{n0}$ and $w_0$ become simultaneously dimensionless 
at $d=2$, which makes $d=2$ the upper critical dimension of the model.
It should be noted, that the total canonical dimension of the field $h$
vanishes for this value of $d$. 

The UV divergences in the Green's functions of the full-scale model
manifest themselves as poles in $\varepsilon=2-d$, and that is why $\varepsilon$
plays the role of the expansion parameter in the RG expansions.

The total canonical dimension of an arbitrary 1-irreducible Green's function
$\Gamma = \langle\Phi \cdots \Phi \rangle_{\rm 1-ir}$ with $\Phi=\{h,h',v\}$
in the frequency--momentum representation is given by the relation:
\begin{equation}
d_{\Gamma}=d+2-d_h N_h-d_{h'}N_{h'}- N_{{\bf v}} d_{{\bf v}},
\label{dGamma}
\end{equation}
where $N_h,N_{h'},N_{{\bf v}} $ are the numbers of the corresponding fields entering
into the function $\Gamma$; see, {\it e.g.} \cite{Book3}.

The total dimension $d_{\Gamma}$ in the logarithmic theory ({\it i.e.} at
$\varepsilon=0$) is, in fact, the formal index of the UV divergence:
$\delta_{\Gamma}=d_{\Gamma}|_{\varepsilon=0}$.
The superficial UV divergences, whose removal requires counterterms, can be
present only in those functions $\Gamma$ for which $\delta_{\Gamma}$ is
a non-negative integer. The counterterm is a polynomial in frequencies and
momenta of degree $\delta_{\Gamma}$ (given that
$\omega\propto k^2$ is implied).

If a number of external momenta occurs as an overall
factor in all diagrams of a certain Green's function, the real index of
divergence $\delta_{\Gamma}'$ will be smaller than $\delta_{\Gamma}$ by
the corresponding number. This happens in our
model:  the derivative at the vertex 
$h'\partial_{\parallel}^2 V(h)$ can be moved onto the field  $h'$ via integration by parts. The derivative in the vertex $-h'(v\partial_{\parallel})h$
can be placed, at will, on $h$ or on $h'$.
 This means that any appearance
of $h'$ in some function $\Gamma$ gives either an external momentum or a square of it, and $\delta_{\Gamma}'$ is either equal to
$\delta_{\Gamma} - N_{h'}$ or $\delta_{\Gamma} - 2N_{h'}$. Moreover, $h'$ can appear
in the corresponding counterterm only in the form of derivative $\partial_{\parallel} h'$.

From table~\ref{table1} and the expression (\ref{dGamma}) one obtains:
\begin{equation}
\delta_{\Gamma}'= \delta_{\Gamma} - 2N_{h'}=4 - 4N_{h'}-N_{{\bf v}},
\label{IndeX2}
\end{equation}
or
\begin{equation}
\delta_{\Gamma}'= \delta_{\Gamma} - N_{h'}=4 - 3N_{h'}-N_{{\bf v}}.
\label{IndeX}
\end{equation}

As all the 1-irreducible Green's functions without the
response fields vanish identically in dynamical models (their diagrams always involve closed
circuits of retarded lines; see, {\it e.g.} \cite{Book3}), it is sufficient to consider only the case $N_{h'}>0$. 

Straightforward analysis of the expression (\ref{IndeX2}), (\ref{IndeX}) shows that
superficial UV divergences can be present only in the
1-irreducible functions of the form $\langle  h'h\dots h \rangle_{1-ir} $ with 
the counter-term $(\partial_{\parallel}^2 h')h^n$ (for any $n\geq 1$). Indeed, all the other 
counter-terms ({\it e.g.} $h'h'$, $h'\partial_{\bot}^2 h$, $h'\partial_t h^n$, and hence from Galilean symmetry $h' (v\partial_{\parallel})h^n$) are not needed 
as the corresponding 1-irreducible functions are finite.

The model is multiplicatively renormalizable because all the terms $(\partial_{\parallel}^2 h')h^n$ are present
in the action (\ref{act1}).
The renormalized action  can be written in the form:
\begin{equation}
{\cal S}_R (\Phi)=h'h'+h'\left\{-\partial_{t}h- v\partial_{\parallel}h+\nu_{\bot} 
\partial_{\bot}^{2} h + Z_{\parallel}\nu_{\parallel}\,
\partial_{\parallel}^{2} h +
\partial_{\parallel}^{2}\sum_{n=2}^{\infty}\frac{Z_n\lambda_{n} h^{n}}{n!}\right\}
+{\cal S}_{\boldsymbol{v}}.
\label{RenAct}
\end{equation}

Here $\nu_{\parallel}$ and $\lambda_n$ are renormalized analogs of the bare
parameters (those with subscript ``o''). ${\cal S}_{\boldsymbol{v}}$ does not require renormalization -- there is no corresponding counter-term -- and the same is true for $\nu_{\bot}$, {\it i.e.} $\nu_{\bot}=\nu_{\bot0}$. 

The renormalization constants $Z_{\parallel}$, $Z_w$, and $Z_n$ depend only on the completely
dimensionless parameters $g_n$ and $w$ and absorb the poles
in $\varepsilon$ and $\xi$. The bare charges $w_0$, $g_0=\{g_{n0}\}$ and bare parameters $\lambda_{n0}$, completely 
dimensionless renormalized charges $w$, $g=\{g_{n}\}$ ($n=2,3,\dots$) and renormalized parameters $\nu_{\parallel}$, $B$, $\lambda_n$ are related as follows:
\begin{equation}
\lambda_{n0}=g_{n0} \nu_{\parallel 0}^{(n+3)/4}\nu_{\bot 0}^{(n-1)/4}, \quad
\lambda_{n}=g_{n} \nu_{\parallel}^{(n+3)/4}\nu_{\bot}^{(n-1)/4}\mu^{\varepsilon(n-1)/2} \label{I61}
\end{equation}
\begin{equation}
B_0=\nu_{\parallel 0} w_0 \quad B = \nu_{\parallel}w\mu^{\xi}.
\label{I6}
\end{equation}
Here the renormalization mass $\mu$ is an additional parameter of the renormalized theory; its canonical 
dimensions
are shown in table~\ref{table1}. 

The renormalized action (\ref{RenAct}) is obtained from the original
one (\ref{act1}) by the renormalization of the parametrs (the renormalization of the fields $h,h',v$ and parameter $\nu_{\bot}$ is not required):
\begin{equation}
\nu_{\parallel 0} =\nu_{\parallel} Z_{\parallel}, \quad
g_{n0}=\mu^{\varepsilon(n-1)/2} g_{n}Z_{g_{n}}, \quad \lambda_{n0} = \lambda_n Z_n, \quad  w_0=wZ_w \mu^{\xi}.
\label{I7}
\end{equation}

The renormalization constants in Eqs. (\ref{RenAct}) and (\ref{I7})
are related as follows:
 \begin{equation}
Z_{g_{n}}=Z_{n}Z_{\parallel}^{-(n+3)/4}, \quad Z_w Z_{\parallel} = 1.
\label{I8}
\end{equation}

Let us consider an elementary derivation of the RG equations \cite{Zinn,Book3}.
The RG equations are written for the renormalized Green's functions
$G_{R} =\langle \Phi\cdots\Phi\rangle_{R}$. In the present case, however, 
the original (unrenormalized) Green's functions $G$ could be considered instead -- the fields are not renormalized and, therefore, $G(e_0,\dots) = 
G_{R}(e,\mu,\dots)$. Here, 
$e_0=\{w_0, g_{n0}, \nu_{\parallel 0}, \nu_{\bot 0}, \dots \}$ is a full set of
bare parameters and $e=\{w, g_{n}, \nu_{\parallel}, \nu_{\bot}, \dots\}$ are their renormalized
counterparts; the ellipsis stands for the other arguments (times,
coordinates, momenta etc.).

We use $\widetilde{\cal D}_{\mu}$ to denote the differential operation
$\mu\partial_{\mu}|_{e_0}$. When expressed in the renormalized variables
it looks as follows:
\begin{equation}
{\cal D}_{RG}\equiv {\cal D}_{\mu} + \sum_{n=2}^{\infty}\beta_{n}\partial_{g_n} + \beta_{w}\partial_{w} 
- \gamma_{\parallel}{\cal D}_{\nu_{\parallel}},
\label{RG2}
\end{equation}
where ${\cal D}_{x}\equiv x\partial_{x}$ for any variable
$x$. The anomalous dimensions $\gamma$ are defined as
\begin{equation}
\gamma_{F}\equiv \widetilde {\cal D}_{\mu} \ln Z_{F} \quad
{\rm for\ any\ quantity} \ F,
\label{RGF1}
\end{equation}
and the $\beta$ functions for the dimensionless coupling constants $g_n$ and $w$ are
\begin{equation}
\beta_{n} \equiv \widetilde {\cal D}_{\mu} g_n = g_n\,[-\varepsilon(n-1)/2-\gamma_{g_n}], \quad
\beta_{w} \equiv \widetilde {\cal D}_{\mu} w = w\,[-\xi-\gamma_{w}].
\label{betagw}
\end{equation}

\section{One-loop expressions for the counterterm, renormalization
constants and RG functions} \label{sec:Funk}

The model involves infinitely many coupling constants. Despite that, the one-loop
counterterm can be obtained - in an explicit closed form in terms of the function $V(h)$. Let us follow the calculation process.

The 1-irreducible Green's functions of our model correspond to the generating functional $\Gamma_{R}(\Phi)$.  Its expansion in the number $p$ of loops looks as follows:
    \begin{equation} 
    \Gamma_{R}(\Phi)=\sum_{p=0}^{\infty}  \Gamma^{(p)}(\Phi),\
    \Gamma^{(0)}(\Phi) = S_{R} (\Phi).
    \label{W19}
    \end{equation}
The loopless (tree-like) contribution is simply the action; the one-loop contribution can be calculated 
via the following relation, see, {\it e.g.} \cite{Book1}:
    \begin{equation}
    \Gamma^{(1)}(\Phi) = - (1/2) {\rm Tr\ ln} (W/W_{0}),
    \label{W20}
    \end{equation}
where $W$ is a linear operation with the kernel
    \begin{equation}
    W(x,y)=-\delta^{2}S_{R} (\Phi) / \delta\Phi(x)\delta\Phi(y),
    \label{W21}
    \end{equation}
and $W_{0}$ is the similar expression for the free parts of the action. 
Both $W$ and $W_{0}$ are $3\times3$-matrices in the set of the fields $ \Phi= \{h,h',v\}$.

By removing UV divergences in (\ref{W19}) and using the minimal subtraction scheme,
we can find the uniquely determined values for constants $Z$. We put $Z=1$ in (\ref{W20}) in the one-loop approximation. In the loopless contribution we keep leading-order terms in the coupling constants 
$g_n$, $w$ in the constants $Z$. For internal consistency
we suppose that $g_n \simeq g_2^{n-2}$.

The Taylor expansion of the function $V(h)$ is
    \begin{equation}
V(h)=\sum^{\infty}_{n=2} \lambda_{n}h^{n}(x)/n!, \quad
V_{R}(h)=\sum^{\infty}_{n=2} Z_{n} \lambda_{n}h^{n}(x)/n!,
    \label{I14}
    \end{equation}
In the following, we interpret similar objects as functions of a single variable $h(x)$, and $V'$, $V''$, etc., 
as the corresponding derivatives with respect to this variable. Thus, the matrix $W$ 
(under the condition that $Z=1$, $v=0$) can be symbolically represented as
\begin{eqnarray}
W=\pmatrix{-\partial_{\parallel}^{2}h'\cdot V''& L^{T} & -\partial_{\parallel}h' \cr L&-2 & \partial_{\parallel}h\cr h'\partial_{\parallel} & \partial_{\parallel}h & D_{v} \cr} \, 
\label{I15}
\end{eqnarray}
where $D_{v} (k)= 2\pi \delta(k_{\parallel}) B_0\, k_{\bot}^{-d+1-\xi}$ from (\ref{veloc2});
$L\equiv \partial_{t}-\nu_{\parallel}\partial_{\parallel}^{2}-\nu_{\bot}\partial_{\bot}^{2}-\partial_{\parallel}^{2}V'$,
and $L^T\equiv -\partial_{t}-\nu_{\parallel}\partial_{\parallel}^{2}-\nu_{\bot}
\partial_{\bot}^{2}-V'\partial_{\parallel}^{2}$ is the transposed operation.

Only the divergent part of expression (\ref{W19}) is required to calculate the constants $Z$; this part was 
previously established to have the form   
\[ \int\ dx\partial_{\parallel}^{2}h'(x) R(h(x)) \]
with a function $R(h)$ similar to $V(h)$. How can one extract this part? Let us recall the well-known formula: $\delta({\rm Tr\ ln}K)={\rm Tr}(K^{-1}\delta K)$
for any variation $\delta K$. By varying the matrix $W$ by $h'$ we obtain 
\begin{eqnarray}
\int\ dx\partial_{\parallel}^{2}h'(x) R(h(x)) \simeq -\frac{1}{2}(-D^{(hh)}V'' \partial_{\parallel}^{2}h' + D^{(hv)} \partial_{\parallel}h' + D^{(vh)} \partial_{\parallel}h') \equiv  \nonumber \\
\equiv  -\frac{1}{2}\int dx\ (-D^{(hh)}V''(h(x)) \partial_{\parallel}^{2}h'(x) + D^{(hv)} \partial_{\parallel}h'(x) + D^{(vh)} \partial_{\parallel}h'(x)),
\label{I16}
\end{eqnarray}
where $D^{(ii)}=(W^{-1})_{ii}$ at $h',v=0$ (the fields are kept equal to zero in $W^{-1}$ because we do not need the terms with them to extract the divergent part). Due to the way it was constructed, $D^{(hh)}$ is the ordinary propagator 
$\langle hh \rangle$ of the model (\ref{RenAct}) with $Z=1$ and with 
$\nu_{\parallel}\partial_{\parallel}^{2}+\nu_{\bot}\partial_{\bot}^{2}+\partial_{\parallel}^{2}V'$ 
substituted for $\nu_{\parallel}\partial_{\parallel}^{2}+\nu_{\bot}\partial_{\bot}^{2}$.

Another consideration should be taken into account. After $\partial^2_{\parallel}$ is 
moved to the external factor $h'$, only a logarithmically divergent expression remains in the counterterm. 
This means that during calculation of the divergent part of a given 
diagram all the external momenta can be put to zero (IR regularization is ensured by the cutoff). Moreover, we can ignore 
the inhomogeneity of $\partial^2_{\parallel}h'(x)$ and $h(x)$, assuming them to be constants in (\ref{I16}) 
while we select the poles in $\varepsilon$ and $\xi$. Then $D^{(hh)}(x,x)$, $D^{(hv)}(x,x)$, and $D^{(vh)}(x,x)$ can be calculated by going over to the 
momentum-frequency representation: 

\begin{eqnarray}
D^{(hh)}(x,x)=\int\int \frac{d\omega d{\bf k}}{(2\pi)^{d+1}} \,
\frac{2}{\omega^{2}+[\nu_{\parallel}k_{\parallel}^{2}+\nu_{\bot}k_{\bot}^{2}+k_{\parallel}^{2}V']^{2}} =
\nonumber \\ =
\frac{S_d}{(2\pi)^d}\frac{\mu^{-\varepsilon}}{ \varepsilon}\frac{1}{\sqrt{\nu_{\bot}(\nu_{\parallel}+V')}} +\dots , \nonumber \\
D^{(hv)}(x,x) = \partial_{\parallel}h\int\int \frac{d\omega d{\bf k}}{(2\pi)^{d}} \,
\frac{B_0 \delta(k_{\parallel}) }{k_{\bot}^{d-1+\xi}(i\omega+\nu_{\parallel}k_{\parallel}^{2}+\nu_{\bot}k_{\bot}^{2}+k_{\parallel}^{2}V')}, 
\nonumber \\
D^{(vh)}(x,x) = \partial_{\parallel}h\int\int \frac{d\omega d{\bf k}}{(2\pi)^{d}} \,
\frac{B_0 \delta(k_{\parallel}) }{k_{\bot}^{d-1+\xi}(-i\omega+\nu_{\parallel}k_{\parallel}^{2}+\nu_{\bot}k_{\bot}^{2}+k_{\parallel}^{2}V')}, 
\nonumber \\
D^{(hv)}(x,x) + D^{(vh)}(x,x) = \partial_{\parallel}h\frac{S_{d-1}}{(2\pi)^{d-1}}\frac{\mu^{-\xi}}{\xi} B_0 +\dots 
\label{I17}
\end{eqnarray}
where the elipsis stands for the UV-finite part; $S_d$ is the surface area of the unit $d$-dimensional sphere: $S_d=2\pi^{d/2}/\Gamma(d/2)$. 

Substituting (\ref{I16}) and (\ref{I17}) into (\ref{W20}), we obtain the following expression for the divergent part of  
$\Gamma_{1}(\Phi)$: 
\begin{eqnarray}
\Gamma_{1}(\Phi)\sim \frac{S_d}{2(2\pi)^d}\frac{\mu^{-\varepsilon}}{ \varepsilon}\int dx
\frac{V''(h(x))}{ \sqrt{ \nu_{\bot}(\nu_{\parallel}+V'(h(x)))} }\, \partial_{\parallel}^{2}h'(x) - \nonumber \\
- \ \frac{S_{d-1}}{2(2\pi)^{d-1}}B_0\frac{\mu^{-\xi}}{ \xi}\int dx\, \partial_{\parallel}^{2}h'(x)h.
\label{I18}
\end{eqnarray}

The sum of (\ref{I18}) and the loopless contribution in (\ref{W20}) has no poles in $\varepsilon$, $\xi$, or their linear combination (they cancel out). This allows us to find the one-loop contributions of order $1/\varepsilon$ and $1/\xi$ in all constants $Z$.

Let us introduce the representation
\begin{equation}
\frac{V''(h(x))}{\sqrt{ \nu_{\bot}(\nu_{\parallel}+V'(h(x)))}} =\sum^{\infty}_{n=0}
\mu^{\varepsilon(n+1)/2} \nu_{\bot}^{(n-1)/4} \nu_{\parallel}^{(n+3)/4}\frac{r_{n}h^{n}}{n!}
\label{I19}
\end{equation}
for the Taylor expansion of the integrand in (\ref{I18}).

Then $r_{n}$ are completely dimensionless coefficients -- polynomials in the charges $g_n$. 
Combining the above condition for the canceling out of poles in $\varepsilon$, $\xi$ and (\ref{I61})-(\ref{I8}), we obtain
\begin{equation}
Z_{\parallel}=1-\frac{r_{1}S_d}{2(2\pi)^d\varepsilon}+\frac{w S_{d-1}}{2(2\pi)^{d-1}\xi}+\dots, \ \  
Z_{n}=1-\frac{r_{n}}{g_{n}}\frac{S_d}{2(2\pi)^d\varepsilon}+\dots\, .
\label{I20}
\end{equation}

The operation $\widetilde{\cal D}_{\mu}$ in (\ref{betagw}) 
assumes the form
\[ \widetilde{\cal D}_{\mu} = \sum_{n}
\left(\widetilde{\cal D}_{\mu}
g_{n}\right)\partial_{g_{n}} + \beta_{w}\partial_{w}= \sum_{n} \beta_{n}\partial_{g_{n}} + \beta_{w}\partial_{w}, \]
which means that in order to achieve the required accuracy it is sufficient to use only the first terms 
in the $\beta$-functions (\ref{betagw}). 
This yields 

\begin{equation}
\widetilde{\cal D}_{\mu} \simeq -\frac{\varepsilon}{2} {\cal D}_{g} - \xi {\cal D}_{w},
\quad
{\cal D}_{g}= \sum^{\infty}_{n=2} (n-1) {\cal D}_{g_{n}}.
\label{I21}
\end{equation}

Applying this to (\ref{RGF1}), (\ref{betagw}), and (\ref{I20}) we obtain the following expressions for 
the one-loop RG-functions:

\begin{equation}
\gamma_{\parallel}=a{\cal D}_{g} r_{1}/2 - bw, \quad a\equiv \frac{S_d}{2(2\pi)^d}, \quad b\equiv \frac{S_{d-1}}{2(2\pi)^{d-1}};
\nonumber
\end{equation}
\begin{equation}
\beta_w=-\xi w + w\gamma_{\parallel};
\label{I22c}
\end{equation}
\begin{equation}
\beta_{n}=-\varepsilon\frac{n-1}{2}g_{n}+\frac{n+3}{4}g_{n}\gamma_{\parallel}-\frac{a}{2}({\cal D}_{g}
-n+1) r_{n}.
\label{I22b}
\end{equation}

Let us consider the explicit expressions for the first four coefficients $r_n$ [the first term with $r_0$ in (\ref{I19}) contributes nothing to (\ref{I18})]; they could be found from the definitions (\ref{I61}), (\ref{I14}), (\ref{I19}):

\[ r_{1}=g_{3}-\frac{1}{2}g_{2}^{2},\quad r_{2}=g_{4}-\frac{3}{2}g_{2}g_{3}+\frac{3}{4}g_{2}^{3}, \]
\[ r_{3}=g_{5}-2g_{2}g_{4} -\frac{3}{2}g_{3}^{2}+\frac{9}{2}g_{2}^{2}g_{3}-\frac{15}{8}g_{2}^{4}, \]
\[  r_{4}=g_{6}-\frac{5}{2}g_{2}g_{5} +\frac{15}{2} g_{2}^{2} g_{4} -5 g_{3}g_{4}+
\frac{45}{4} g_{2}g_{3}^{2}-\frac{75}{4} g_{2}^{3}g_{3} +\frac{105}{16} g_{2}^{5} , \]

when substituted into (\ref{I22b}) they yield: 

\begin{eqnarray}
\gamma_{\parallel} &=& \frac{a}{2} (2g_{3}-g_{2}^{2}) - bw,
\label{I23a} \\
\beta_w&=&-\xi w + w\frac{a}{2} (2g_{3}-g_{2}^{2})-bw^2, 
\label{I23c} \\
\beta_{2}&=&\left(-\frac{\varepsilon}{2} -\frac{5}{4}bw\right)g_{2}+a\left(-g_{4} +\frac{11}{4} g_{2}g_{3}-\frac{11}{8} g_{2}^{3}\right),
\label{I23bb}\\
\beta_{3}&=&\left(-\varepsilon-\frac{3}{2}bw \right)g_{3}+a\left(-g_{5}+2g_{2}g_{4}+3 g_{3}^{2}-\frac{21}{4} 
g_{2}^{2}g_{3} +\frac{15}{8}  g_{2}^{4}\right).
\label{I23b}
\end{eqnarray}

(We recall that we have to admit $g_{n} \sim g_{2}^{(n-1)}$ for the sake of consistency of the approximation.) These two examples -- $\beta_{2}$ and $\beta_{3}$ -- give us the general form of the functions (\ref{I22b}).

\section{Attractors and critical exponents} \label{sec:Att}

Let us turn to the complete system (\ref{I22c}), (\ref{I22b}) of the $\beta$-functions. The fixed points of RG equations 
 can be found from the requirement that  $\beta_{w}(w_{*},g_{*})=0$, $\beta_{n}(w_{*},g_{*})=0$, $n=2,3,\dots$. The first equation $\beta_{w}(w_{*},g_{*})=0$ has two solutions: $w_{*}^{(1)}=0$ and $w_{*}^{(2)}=(-\xi+a(2g_{3*}-g_{2*}^{2})/2)/b$. The explicit form of the $\beta$-functions (\ref{I23c}), (\ref{I23bb}), (\ref{I23b}) shows that we can choose the coordinates 
$g_{2*}$, and $g_{3*}$ arbitrarily, while all the other $g_{n*}$ with $n\ge4$ are then  uniquely determined 
from the equations $\beta_{k}(g_{*})=0$, $k\ge3$. Instead of a set of a fixed points in the infinite-dimensional space of the 
couplings $w,g\equiv \{ w,g_{n} \}$, the RG-equation (\ref{RG2}) has two two-dimensional surfaces of fixed points, parametrized by the values of $g_{2*}$, and $g_{3*}$, with either $w_{*}^{(1)}=0$ or $w_{*}^{(2)}=(-\xi+a(2g_{3*}-g_{2*}^{2})/2)/b$. The former surface corresponds to IR asymptotic regime, where turbulent mixing is irrelevant; it coincides with the surface obtained in \cite{US}.

In general, it is difficult to establish the character of these fixed points. According to the general rule 
\cite{Zinn}, a point $w^*,g^*\equiv \{ w^*, g_{n}^*\}$ is IR stable if the real parts of all the eigen-numbers of the matrix 
$\omega_{kl}=\partial\beta_{k}/\partial g_{l}|_{w_{*},g_{*}} $ (where $\omega_{11}=\partial\beta_{w}/\partial w|_{w_{*},g_{*}} $) are strictly positive. The requirement that  
all the diagonal elements $\omega_{kk}$ be positive is the necessary  condition for IR-stabihty. 
Equations (\ref{I22c}), (\ref{I22b}) yield these elements for all values of $k$: 

\[ \omega_{11}=-\xi+\frac{a}{2} \left[2g_{3*}-g_{2*}^{2}\right]-2bw_*,\]
\[ \omega_{22}=-\frac{\varepsilon}{2}+a\left[\frac{11}{4} g_{3*}-\frac{33}{8} g_{2*}^{2}\right]-\frac{5}{4}bw_*,\]
\[\omega_{33}=-\varepsilon+a \left[6g_{3*}-\frac{21}{4} g_{2*}^{2}\right]-\frac{3}{2}bw_*, \]
and for $n\ge4$ we have
\[ \omega_{nn}= -\varepsilon\frac{n-1}{2}+a \frac{(n+1)^2+2}{4} g_{3*}-a\frac{n(3n+4)+3}{8} g_{2*}^{2}-\frac{n+3}{4}bw_*. \]

For the most realistic values of $\varepsilon$ and $\xi$ ($0$ and $2$), regions where all these quantities are positive exist. However, this is just a necessary condition; still, we can assume that the surfaces of fixed points 
$w_{*},g_{*}$ contain regions of IR stability. If this is indeed so, the model exhibits IR scaling with 
nonuniversal critical dimensions, ({\it i.e.} they depend on the the parameters $g_{2*}$, and $g_{3*}$). 

In dynamic models of the type (\ref{act1}) the critical exponents $\Delta_F$ of an arbitrary quantity
 $F$ (a field or a parameter) is given by the following expression (for detailed explanation see, {e.g.} \cite{Alexa}):
\begin{eqnarray}
\Delta_{F} = d^{\bot}_{F} + d^{\parallel}_{F}\Delta_{\parallel} +d^{\omega}_{F}\Delta_{\omega}  +
\gamma_{F}^{*}, \quad \Delta_{w} =2-\gamma_{\bot}^{*}, \quad \Delta_{\parallel}= 1 + \gamma_{\parallel}^{*}/2.
\label{Dimension}
\end{eqnarray}

For $F= h$ we have $\gamma_{h}^{*}=0$ and $\gamma_{\bot}^{*}=0$ (the fields and the parameter 
$\nu_{\bot}$ are not renormalized). Relations (\ref{Dimension}) together with the table~1 yield the exact result 
$2\Delta_{h}=d-1+\Delta_{\parallel}-\Delta_{\omega}$; from (\ref{I23a}) we find
that $\Delta_{\parallel}=1+a (2g_{3*}-g_{2*}^{2})/4-bw_*/2$, $\Delta_{h}=a (2g_{3*}-g_{2*}^{2})/8-bw_*/4$  in the one-loop approximation. 

\section{Conclusion} \label{sec:Conc}

We applied the standard field
theoretic RG to the model of landscape erosion (proposed in \cite{Pastor1, Pastor2} and modified in \cite{US}) subjected to advection by anisotropic velocity ensemble \cite{AM}. It turned out that the model could be reformulated as a multiplicatively renormalizable field theoretic model
with an infinite set of independent renormalization constants (thus, infinite set of coupling constants). Despite this fact, it appears possible to derive the one-loop counterterm employing the 
method earlier proposed in  \cite{US,AV} for an isotropic model
of surface roughening. The method yields two two-dimensional surfaces of fixed points; one of them corresponds to IR asymptotic regime, where turbulent mixing is irrelevant, and coincides with the surface obtained in \cite{US}. 

These surfaces of fixed points are likely to 
contain IR attractive region(s). If that is the case, then the model exhibits scaling behaviour. 
The corresponding scaling exponents turn out to be nonuniversal because of their dependence
on the coordinates of specific fixed point on the surfaces. Nonetheless, they satisfy certain exact universal relation, that, in principle, can be tested experimentally.

From a more theoretical point of view, it is desirable to write down the RG equations and to find the attractors directly in terms of the function $V(h)$, so that, instead of infinitely many $\beta$ functions for the couplings, we would have the only functional $\beta(V)$  with the only functional argument $V(h)$; see the discussion in \cite{Dima} for a general case. This work remains for the future and is partly in progress.

\section*{Acknowledgments}
The authors thank L. Ts. Adzhemyan, M. Hnatich, J. Honkonen and M. Yu. Nalimov for discussion.
The authors thank the Organizers of the International Conference
``Quarks 2016'' for the opportunity to present the
results of their research.
The authors also acknowledge the Saint-Petersburg State University for
research grant 11.38.185.2014. One of the authors (P.K.) was also supported by the RFBR research grant 16-32-00086.

\end{document}